# Accurate energy-size dependence of excitonic transitions in semiconductor nanocrystals and nanoplatelets using a phase jump approach

**Arjen T. Dijksman[1], and Benoit Dubertret[*,1]**

[1] Ecole Supérieure de Physique et de Chimie Industrielles, ParisTech, 10 rue Vauquelin, 75005 Paris, France

* Corresponding author: e-mail benoit.dubertret@espci.fr, Phone: +33 1 40 79 45 92, Fax: +33 1 40 79 45 37

Accurate energy-size dependence of excitonic transitions in semiconductor nanocrystals in the strong confinement regime using classical theoretical approaches such as effective mass approximation, tight binding, or empirical pseudo-potential is difficult. We propose a simple empirical expression with three fitting parameters that accurately relates the size dependence of most known excitonic transitions in CdSe and in InAs nanocrystals. We show that this empirical expression can be deduced from a phase jump approach if the charge carriers are considered to travel on the atomic lattice of the nanocrystal and gain energy upon bouncing at the nanoparticle boundaries. This empirical expression is also tested on the atomically flat CdSe nanoplatelets without any adjustment of the parameters obtained with the CdSe spherical nanocrystals, and provides an estimation of the CdSe nanoplatelets thickness that matches exactly the experimental observations. These results suggest that a phase shift approach could be useful to describe the electronic transitions in semiconductor nanocrystals.

**1 Introduction** A growing appreciation of the exceptional physical properties of colloidal semiconductor nanocrystals has motivated extensive studies of these materials since their discovery in the early 1980s [1–3]. Technological know-how, optimized colloidal syntheses, new experimental data and theoretical modeling steadily increased our understanding of the effects of the confinement of charge carriers in colloidal semiconductor nanocrystals, also called quantum dots (QDs) [4]. They constitute without doubt a very promising research field for the design of innovative materials. Several theoretical approaches have been used to describe the physics of the electronic transitions in QDs. They include essentially effective mass (EM) approximation [5], tight-binding (TB) [6–8] and empirical pseudo-potential [9–11] approaches. These different approaches have led to a much better understanding of the confinement effect in the QDs, but in the case of the excitonic transition dependence upon the nanocrystal size, their predictions remain limited. For example, in the case of effective mass approximation, discrepancies between theory and experimental data are observed [12], due to our incomplete knowledge of such characteristics as finite barrier, inter band coupling, exchange interaction and size dependence of the dielectric constant. For practical applications the size of QDs is deduced from their absorption spectra using an empirical polynomial fit [13,14].

In this paper, we present an alternative way to investigate the absorption spectra of semiconductor nanocrystals in the strong confinement regime, which leads to empirical expressions for the energy-size dependence of excitonic transitions. It includes a phase jump term in the solutions of the Schrödinger equation. The model is probed against high quality experimental data of CdSe and InAs QDs of the literature and shows good agreement in the strong confinement range. It is further verified and specified on semiconductor nanoplatelets, which provide a good benchmark for the comparison with different approaches.

**2 Theoretical basis** We propose the following size-dependent expression for the energy of the $i^{th}$ excitonic transition:

$$E^i(a) = E_0^i \frac{a+\gamma_\phi^i}{a+\gamma_t^i} \quad (1)$$

where $E_0^i$ is a constant reference energy gap for the $i^{th}$ excitonic transition and $\gamma_\phi^i$ and $\gamma_t^i$ are respectively metric phase and time delay parameters for the same transition. Their physical interpretation will be discussed later.

Figure 1 shows the high quality experimental data of the first excitonic transition of CdSe QDs published by Norris *et al.* [12] and their fit with Eq. (1). The fit is very good (coefficient of determination $R$=0.9997) and suggests that Eq. (1) could be used for size determination, equivalently to empirical polynomial fitting formulas of higher order [13,14]. Equation (1) can also be used to fit the higher energy transitions of CdSe (see Fig. 2). For each transition, the set of parameters has to be adjusted, but the fit with the experimental data is strikingly good. Equation (1) can also be used to fit the InAs excitonic transitions reported by Banin *et al.* [15] with an accuracy equivalent to that of CdSe (see Fig. 3). This suggests that Eq. (1) can be used with success for various materials. In the following section we provide a framework that explains how we obtained Eq. (1).

(2). The output values for parameters $E_0^i$ (eV), $\gamma_\phi^i$ (nm), $\gamma_t^i$ (nm) and the coefficient of determination $R^2$ are given in Table 1. Dotted lines denote weak transitions. (Experimental data, courtesy of D. J. Norris) [12].

**Table 1** Parameters for the CdSe optical transitions of Fig. 2.

| Transition | $E_0$ (eV) | $\gamma_\phi$ (nm) | $\gamma_t$ (nm) | $R^2$ |
|---|---|---|---|---|
| (a) | 1.40 | 2.34 | 0.60 | 0.9997 |
| (b) | 1.38 | 2.26 | 0.37 | 0.9993 |
| (c) | 1.41 | 2.46 | 0.42 | 0.9999 |
| (d) | 1.16 | 4.76 | 0.83 | 0.998 |
| (e) | 1.51 | 2.40 | 0.30 | 0.995 |
| (f) | 0.20 | 58.0 | 1.97 | 0.996 |
| (g) | 1.08 | 13.8 | 1.42 | 0.997 |
| (h) | 1.25 | 9.90 | 0.70 | 0.995 |
| (i) | 1.16 | 12.2 | 0.70 | 0.990 |
| (j) | 2.34 | -2.63 | -3.81 | 0.989 |

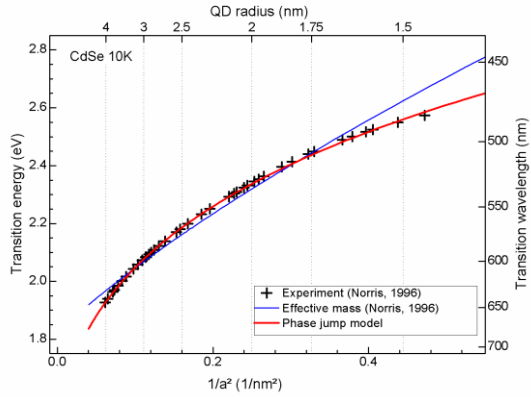

**Figure 1** Energy from the first excited state in spherical CdSe nanocrystals as a function of $1/a^2$. Equation (1) is fitted with a coefficient of determination 0.9997 for parameters $E_0$=1.40 eV, $\gamma_\phi$ = 2.34 nm, $\gamma_t$ = 0.60 nm. (Experimental data and effective mass approximation from Ref. 12, courtesy of D. J. Norris).

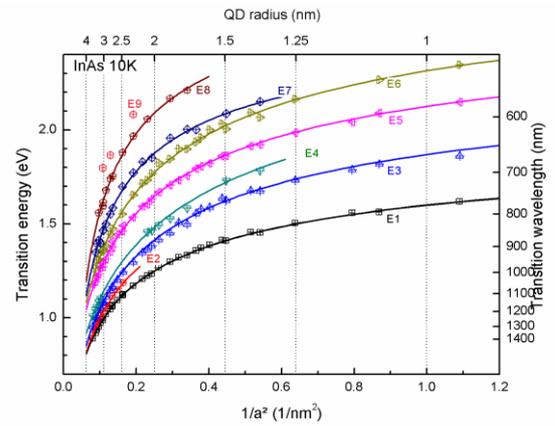

**Figure 3** Energy from excitonic transitions E1 to E9 of InAs, plotted as a function of $1/a^2$, with best fit functions determined from Eq. (1). The output values for parameters $E_0^i$ (eV), $\gamma_\phi^i$ (nm), $\gamma_t^i$ (nm) and the coefficient of determination $R^2$ are given in Table 2. (Experimental data for $a > 1.3$ nm from Ref. 15, for $a < 1.3$ nm, courtesy of U. Banin).

**Table 2** Parameters for the InAs optical transitions of Fig. 3.

| Transition | $E_0$ (eV) | $\gamma_\phi$ (nm) | $\gamma_t$ (nm) | $R^2$ |
|---|---|---|---|---|
| E1 | -0.55 | -16.0 | 4.16 | 0.999 |
| E2 | -0.84 | -12.1 | 4.36 | 0.998 |
| E3 | -0.80 | -12.4 | 3.88 | 0.998 |
| E4 | -0.28 | -24.4 | 2.23 | 0.999 |
| E5 | -0.80 | -14.6 | 4.12 | 0.999 |
| E6 | -1.13 | -12.0 | 4.37 | 0.998 |
| E7 | -7.58 | -7.69 | 21.0 | 0.997 |
| E8 | -31.0 | -6.86 | 70.1 | 0.996 |

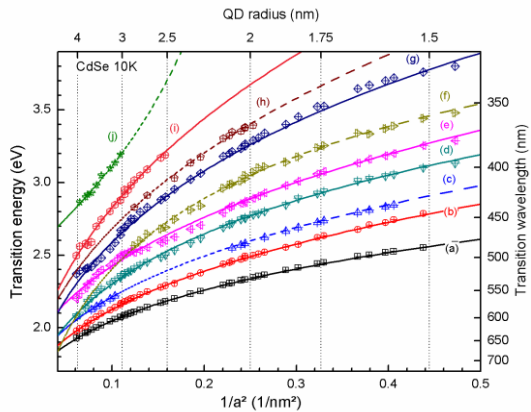

**Figure 2** Energy from excitonic transitions (a) to (j) of CdSe (Ref. 12), versus $1/a^2$, with best fit functions determined from Eq.

**3 Phase jump approach** We will now propose a semi-empirical deduction of Eq. (1).

We assume the electron to be a localized particle which is moving through the crystalline structure. As an approximation, we will consider the "classical" path where it travels in a right line from atom to atom. Let us represent the electron's phase by a rotating vector. As the electron's phase $\phi(t)$ is progressing, the direction of the vector is changing, equivalently to a quantum stopwatch [16], see Fig. 4. Its angular velocity $\omega(t) = d\phi/dt$ determines its energy $E(t) = \hbar\omega(t)$. In a steady state, $\phi(t) = \omega_a t$, and the action $A$ advances linearly with time at a rate equal to the energy $E_a$ of the electron:

$$E_a = \frac{A(t_2)-A(t_1)}{t_2-t_1} = \frac{\hbar\Delta\phi}{\Delta t} = \hbar\omega_a \quad (2)$$

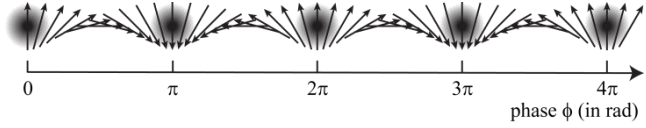

**Figure 4** Illustration of a rotating vector representing the phase progression of an electron travelling from left to right on its path from atom to atom. The phase progresses at $\pi$ rad per atom-atom length.

Since the electron moves from atom to atom, Eq. (2) can also be written as $E_a = \hbar\phi_a/t_a$, where $\phi_a$ is the phase shift and $t_a$ is the delay needed when the electron travels over a distance $x_a$ between two consecutive identical atoms.

We suppose that the particle's phase evolves in phase with the global standing wave solutions of the nanocrystal. One could speak of phase matching between a pilot wave and a guided quantum particle [17,18]. In order to stay in phase with its environment, the rotating vector of the guided electron must continuously adapt its angular velocity.

We now analyze the influence of an abrupt interface. When the electron reaches an interface, it may bounce or reflect on it, if not trapped. Upon this reflection, the electron's rotating vector must readapt its phase in order to match phases with the standing wave profile in the new direction. If this phase jump $\phi_s$ is performed during a time interval $t_s$, the angular velocity of the rotating vector is temporarily changed. In a nanoparticle with small dimensions, the electron bounces with high frequencies on the nanoparticle surface, producing phase jumps at angular velocity $\omega_s = \phi_s/t_s$.

In a steady state, the rotating vector's angular velocity oscillates between two values: the steady-state bulk value $\omega_a = \phi_a/t_a$ and the phase jump value $\omega_s = \phi_s/t_s$ at the surface. The action carried out by an electron that progressed over a distance $nx_a$ along nearest identical atoms and bounced on $p$ interfaces is:

$$A = (n\phi_a + p\phi_s)\hbar \quad (3)$$

and the energy associated to this path is:

$$E(n,p) = \frac{n\phi_a + p\phi_s}{nt_a + pt_s}\hbar. \quad (4)$$

If this phase jump is periodical, such that in between two consecutive bounces on interfaces, the particle travels over a distance $nx_a$, the average energy is:

$$E(n) = E_a \frac{n+\rho_\phi}{n+\rho_t} \quad (5)$$

where $\rho_\phi = \phi_s/\phi_a$ and $\rho_t = t_s/t_a$ are the phase jump and time delay ratios at the surface with respect to the regular phase and time progression between two nearest identical atoms. Equation (5) and (1) are mathematically equivalent. In Eq. (5), the number $n$ of atomic distances over which the electron progresses in between two reflections is proportional to the dimension of the nanoparticles.

In this model of strong confinement, we suppose that the energetical transitions of both electrons and holes can be described with Eq. (5), and that both charge carriers have the same time delay ratio $\rho_t$. It results that the electron and hole contributions can be summed up and Eq. (5) can be used for all excitonic transitions under strong confinement. This assumption holds if the masses of the charge carriers cancel out in the ratio $\rho_t$.

**4 Discussion** Equation (5) has three parameters that we now discuss.

$E_a$ in Eq. (5) corresponds to the energy of the transitions for $n \to \infty$, i.e. for the bulk material. For the first excitonic transition, this should correspond to the energy gap $E_g$. However, the fitted reference energies $E_0^i$ obtained by fits of Eq. (1) are considerably smaller than the bulk bandgap energy $E_g$ (except for transition (j)). This discrepancy means that the phase jump model can only be applied in the strong confinement regime.

The two other parameters in Eq. (5) are related to the phase jump both at the interfaces and between atoms, as well as the time necessary for these phase jumps. If this approach is valid, these parameters should be identical in identically structured materials whatever the shape of the nanocrystal, provided that the effect of surface chemistry on the phase jumps is identical. We tested this effect with CdSe nanoparticles of different shapes.

Fortunately, CdSe nanocrystals have been synthesized in different shapes including the spheres [19,20] and more recently the nanoplatelets [21]. The advantage of the nanoplatelets is that they have been shown to be atomically flat with a thickness that is controlled with atomic preci-sion [21]. The crystal structure as well as the crystal orientation of the nanoplatelet facets have been documented, and they have been shown to have pure 1D confinement. They are thus a very interesting object to test this phase jump approach without different fitting parameters.

In a nanoplatelet, we hypothesize that the charge carrier bounces back and forth between the two extended (001) surfaces of the nanoplatelet, experiencing a phase jump at each reflection, in analogy to a photon in a Fabry-Perot resonator, as presented in Fig. 5. The path we propose for the first transition is the shortest path for a particle moving from Cd atom to Cd atom using the nearest atom at each jump. If we transpose this path to a spherical QD, with identical phase jump and time delay ratios, the charge carriers should circle inside the sphere, bouncing 4 times for a complete cycle, see Fig. 6.

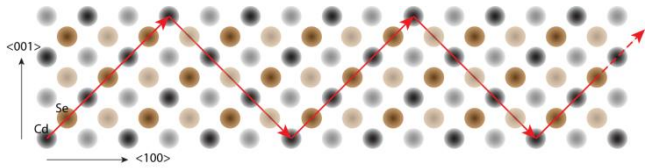

**Figure 5** Schematic side view of a zinc blende CdSe 3 monolayer thick nanoplatelet. The excited electron starts along <101> direction. After a phase progression $3\phi_a$, it reflects at the (001) upper surface with phase jump $\phi_s$, continues along <1,0,-1> direction, reflects again, *etc*.

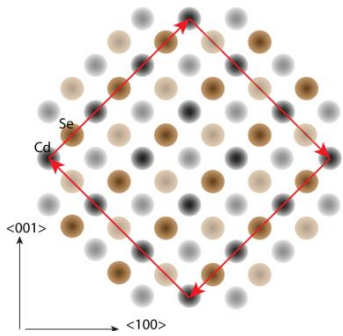

**Figure 6** Schematic side view of a zinc blende CdSe 3 monolayer radius QD. Electron starts along <101> direction. After a phase progression $3\phi_a$, it reflects at (001) upper surface with phase jump $\phi_s$, continues along <1,0,-1> direction, reflects again, continues along <-1,0,-1> direction, reflects again, *etc*.

If such a path holds for spherical QDs, a nanoplatelet with a thickness of $n$ monolayers should have the same transition as a spherical QD with a radius of $n$ monolayers, provided that the two systems have the same nearest neighbour distance, the same composition (and thus the same $\phi_a$) and similar surface chemistries (and thus the same $\phi_s$).

In Fig. 7, we plot the experimental room temperature energy of the first excitonic CdSe transition for nanoplatelets *vs.* their thickness in monolayers [22] and for QDs *vs.* their radius in shell monolayers [20]. The results coincide very well and can both be fit with the same parameters using Eq. (5). In addition, the NPL thicknesses we deduce from this model are exactly the ones measured on CdSe NPLs using HRTEM [23]. The experimental observation that a nanoplatelet with a thickness of $n$ monolayers has its first absorption feature at the same wavelength as a spherical quantum dot with a radius corresponding to $n$ CdSe monolayers is thus confirmed by this approach. Of course, this result only holds for the hypothesized path along the nearest cation-cation direction. Other transitions may result from other paths, phase jumps or time delays.

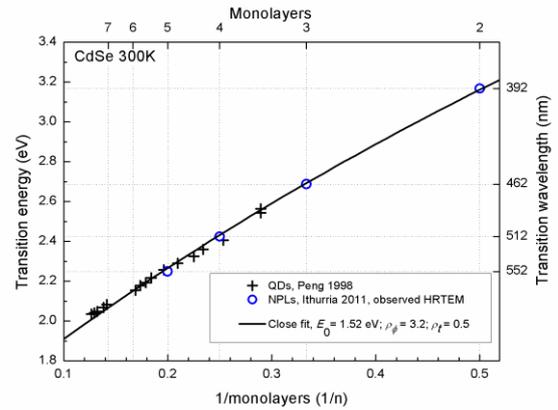

**Figure 7** Comparison between first absorption peak of CdSe QDs (crosses) [20] respectively CdSe nanoplatelets (NPLs) (open figures) [22] *vs.* their radii respectively thicknesses expressed in terms of monolayers (the QD literature sizes in nm are divided by 0.304, which corresponds to one monolayer in the zinc-blende <100> direction). The solid line represents a close fit with $E_0 = 1.52$ eV, $\rho_\phi = 3.2$, $\rho_t = 0.5$.

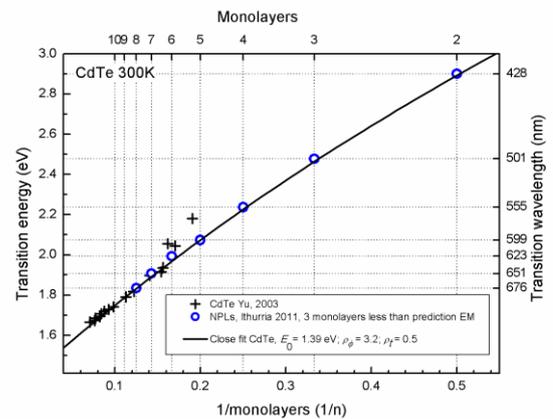

**Figure 8** Comparison between first absorption peak of CdTe QDs (crosses) [13] respectively CdSe nanoplatelets (NPLs) (open figures) [22] *vs.* their radii respectively thicknesses expressed in terms of monolayers (the QD literature sizes in nm read from the graph of Ref. 13 are divided by 0.324, which corresponds to one

monolayer in the zinc-blende <100> direction). The solid line represents a close fit with $E_0$=1.39 eV, $\rho_\phi$ = 3.2, $\rho_t$=0.5.

We have also compared the energy of the first transition of CdTe spherical QDs [13] with CdTe nanoplatelets [22]. We first fit the CdTe QDs size dependence using Eq. (5), leaving out the 3 smallest sizes. We then use their parameters without any change to deduce the thicknesses of CdTe nanoplatelets. The results are reported in Fig. 8. While the exact thicknesses of CdTe nanoplatelets have not been obtained by HRTEM yet, we observe a difference of 3 monolayers in thickness estimation compared to EM approximation. Interestingly, the fitting of Eq. (5) for the first transition of CdSe and CdTe QDs shows that $\rho_\phi$ and $\rho_t$ are identical for the two materials. Only $E_0$ is different.

**5 Conclusion** We propose a simple expression that relates the electronic transitions of semiconductor nanocrystals with their size. This expression has three parameters that are used to fit with a very good accuracy all the excitonic transitions of CdSe and InAs spherical nanocrystals as well as the excitonic transitions of atomically flat CdSe and CdTe nanoplatelets. We show that this simple expression can be deduced from a rotating vector representation of the charge carriers in the nanocrystal travelling from atoms to atoms and bouncing on the crystal facets. This simple phase jump model, although semi-empirical, provides a very precise relationship between the size of a nanocrystal and its electronic transitions. It provides a very simple description of all the electronic transitions of the nanocrystals we have investigated, that seems to be accurate at least in the strong confinement regime.

It can be used to predict successfully the transitions of some nanocrystals. For example, we show that the parameters used to fit the excitonic transitions of spherical CdSe nanocrystals can be used without adjustment to fit the excitonic transitions of CdSe nanoplatelets in the pure 1D confinement regime, as expected in this model.

This approach considers semiclassical periodic paths in closed systems. It shows some similarities with other semiclassical treatments, like chaotic quantum systems considered with Gutzwiller or de Broglie-Bohm periodic trajectories [24–26]. It would therefore be interesting to study the application of these semiclassical methods to the QDs and NPLs we have investigated in this paper. We hope that the good agreement of this phase jump model with experimental observations will foster further investigation and identification with other approaches.

**Acknowledgements** The authors thank C. Ciuti, A. Bramati, M. Tessier and T. Pons for stimulating discussions. A.T.D. acknowledges funding by the Agence Nationale de la Recherche.